\documentclass[reprint,amsmath,amssymb,superscriptaddress,aps,prfluids,onecolumn,nofootinbib]{revtex4-2}
\usepackage[utf8]{inputenc}
\usepackage[top=25mm, bottom=25mm, left=25mm, right=25mm]{geometry}
\usepackage{graphicx}
\usepackage{epstopdf}
\usepackage{amssymb}
\usepackage{amsmath}
\usepackage[hidelinks]{hyperref}
\usepackage{float}
\usepackage[usenames, dvipsnames]{xcolor}
\usepackage{soul}
\usepackage[shortlabels]{enumitem}
\usepackage[font=small]{caption}
\usepackage{siunitx}
\usepackage{bm}
\usepackage{bbold}
\usepackage[bbgreekl]{mathbbol}

\DeclareSymbolFontAlphabet{\mathbb}{AMSb}

\begin{document}


\title{Viscoelasticity reduces the droplet size in mucosalivary film fragmentation during intense respiratory events}
\def\thefootnote{$\S$}\footnotetext{These authors contributed equally to this work.}
\author{Mogeng Li$^\S$}
\affiliation{Physics of Fluids Department, Faculty of Science and Technology, University of Twente, 7500 AE Enschede, The Netherlands}
\affiliation{School of Aerospace, Mechanical, and Mechatronic Engineering, University of Sydney, Sydney NSW 2008, Australia}
\author{Youssef Saade$^\S$}
\affiliation{Physics of Fluids Department, Faculty of Science and Technology, University of Twente, 7500 AE Enschede, The Netherlands}
\affiliation{Canon Production Printing Netherlands B.V., 5900 MA Venlo, The Netherlands}
\author{St\'{e}phane Zaleski}
\affiliation{Sorbonne Universit\'{e} and CNRS, Institut Jean Le Rond d'Alembert UMR 7190, F-75005 Paris, France}
\affiliation{Institut Universitaire de France, Paris, France}
\author{Uddalok Sen}
\email{uddalok.sen@wur.nl}
\affiliation{Physics of Fluids Department, Faculty of Science and Technology, University of Twente, 7500 AE Enschede, The Netherlands}
\affiliation{Physical Chemistry and Soft Matter group, Wageningen University and Research, 6708 WE Wageningen, The Netherlands}
\author{Pallav Kant}
\affiliation{Physics of Fluids Department, Faculty of Science and Technology, University of Twente, 7500 AE Enschede, The Netherlands}
\affiliation{Department of Mechanical and Aerospace Engineering, University of Manchester, M13 9PL Manchester, United Kingdom}
\author{Detlef Lohse}
\email{d.lohse@utwente.nl}
\affiliation{Physics of Fluids Department, Faculty of Science and Technology, University of Twente, 7500 AE Enschede, The Netherlands}
\affiliation{Max Planck Institute for Dynamics and Self-Organization, 37077 G\"{o}ttingen, Germany}

\date{\today}

\begin{abstract}
    We examine the fundamental fluid dynamical mechanisms dictating the generation of bioaerosols in the human trachea during intense respiratory events such as coughing and sneezing, with an emphasis on the role played by the mucosalivary fluid viscoelasticity. An experimental investigation of the shear-induced fragmentation of a mucosalivary-mimetic fluid in a confined geometry reveals that viscoelastic liquids undergo atomization in a manner akin to Newtonian liquids -- via the formation of bag-like structures -- which ultimately rupture through the appearance of retracting holes on the bag surface. Droplets are produced via the unstable retraction of liquid rims bounding these holes. However, in comparison to Newtonian liquids, viscoelastic bags inflate to larger sizes -- implying thinner sheets and, consequently smaller droplets upon rupture. Numerical simulations support that the smaller droplets can be attributed to the thinner sheets, with a more uniform thickness, for viscoelastic bags prior to rupture. Hence, we highlight the role of the viscoelasticity in determining the thickness of the intermediate bag-like structures, which, in turn, govern the droplet size distribution of the expelled aerosol. 
\end{abstract}

\maketitle


\section{Introduction} \label{sec:intro}

Respiratory droplets (bioaerosols) are expelled from the mouth and nose during activities such as speaking, coughing, and sneezing \citep{xie-2009-jrsi, bourouiba-2014-jfm, scharfman-2016-expfluids, abkarian-2020-pnas, bourouiba-2021-arfm, bourrianne-2021-prf, bourrianne-2022-prappl, nord-2024-jaerosolsci}. These droplets act as pathogen carriers, and play a significant role in the transmission of respiratory diseases \citep{wells-1936-jama, jones-2015-joccupenvironmed, fiegel-2006-drugdiscovtoday, bourouiba-2016-nejm, abkarian-2020-prf, bourouiba-2021-arfm, pohlker-2023-rmp, gidreta-2023-jfm}, as evident from the recent COVID-19 pandemic \citep{bourouiba-2020-jama, mittal-2020-jfm, somsen-2020-lancetrespirmed, wang-2021-science}. In addition to ventilation \citep{yang-2022-jfm, stockwell-2019-jhospinfect, morawska-2020-environint, morawska-2021-science}, temperature, and humidity of the environment \citep{ng-2021-prf, chong-2021-prl}, the fate of the exhaled droplets obviously depend strongly on their sizes \citep{wells-1934-amjepidemiol, ng-2021-prf, bourouiba-2014-jfm} (diameters typically in the sub-micrometric to millimetric range \citep{duguid-1946-epidemiolinfect, loudon-1967-nature, chao-2009-jaerosolsci}): larger droplets tend to fall ballistically and deposit on surfaces while smaller droplets can be sheltered in the exhaled puff and remain airborne for a longer time and over larger distances \citep{bourouiba-2014-jfm, bourouiba-2016-nejm, bourouiba-2020-jama, bourouiba-2021-arfm, ng-2021-prf, chong-2021-prl}. This, of course, has a strong bearing on the effectiveness of conventional mitigation measures such as social distancing and ventilation. Newer respiratory disease containment strategies focusing on mitigating exhaled bioaerosols can be enabled by directly altering the airway-lining fluid properties \citep{edwards-2004-pnas, fiegel-2006-drugdiscovtoday}. Therefore, a sound understanding of the mechanisms dictating the generation of bioaerosols, and the resulting droplet size distribution, is imperative for improving mitigation strategies. 

Three major modes of droplet generation have been identified in human respiratory tracts \citep{wei-2016-amjinfectcontrol}: (i) droplets can be produced via the rupture of a low surface tension liquid film with the collapsing and opening of bronchioles \citep{johnson-2009-jaerosolmedpulmdrugdeliv}, shear instability in (ii) the trachea \citep{fiegel-2006-drugdiscovtoday, vasudevan-2005-intjengsci, moriarty-1999-jfm} and (iii) the narrow passage at the opening of the mouth \citep{morawska-2005-conf, abkarian-2020-prf}. The bronchiole mode mainly contributes to the production of micrometric or sub-micrometric droplets, and the oral cavity mode is responsible for larger millimetric droplets. In our previous work \citep{kant-2023-prf}, we, both experimentally (using an in-house constructed ``cough machine" \citep{edwards-2004-pnas, king-1985-japplphysiol, clarke-1970-japplphysiol}) and numerically (using the Volume-of-Fluid method \citep{marcotte-2019-prf, pairetti-2021-computfluids, kulkarni-2024-arxiv, shen-2025-jfm}), studied the second mode of droplet production, which is considered to be the most relevant one in the context of coughing or sneezing \citep{evrensel-1993-jbiomedeng, moriarty-1999-jfm, bake-2019-respirres}. A high velocity gas phase flowing over a thin liquid film shears the latter, resulting in the formation of creases, which subsequently inflate into `bags', and eventually disintegrate into droplets \citep{blanchard-1963-progoceanogr, basser-1989-jbiomecheng, moriarty-1999-jfm, bremond-2005-jfm, villermaux-2009-natphys, vledouts-2016-jfm, villermaux-2020-jfm}. For a single droplet exposed to aerodynamic forces (free-falling \citep{villermaux-2009-natphys, jalaal-2012-ijmf} or immersed in a stream of high-speed airflow \citep{hinze-1955-aichej, theofanous-2011-arfm, jackiw-2021-jfm, chandra-2024-prf}), the fragmentation morphology has historically been classified into four regimes \citep{villermaux-2020-jfm}: bag (also known as ``Rayleigh-Taylor piercing"), multimode (sometimes referred to as ``bag and stamen" and ``multibag"), sheet thinning, and catastrophic \citep{guildenbecher-2009-expfluids, zhao-2010-pof}. The fragmentation mode observed in the cough machine setup resembles the bag mode in aerodynamic breakup. This regime is typically characterized by a Weber number ($\mathrm{We}$) range of $11 \lesssim \mathrm{We} \lesssim 18$, where $\mathrm{We}$ denotes the ratio of the inertial-to-surface tension forces \citep{jalaal-2012-ijmf, krzeczkowski-1980-ijmf, hsiang-1992-ijmf, guildenbecher-2009-expfluids}. Similar bag breakup modes have also been observed in other practically-relevant systems, such as a liquid sheet under cross flow \citep{varkevisser-2024-jfm} and wind-generated waves at the air-sea interface \citep{veron-2012-geophysreslett, troitskaya-2017-scirep, villermaux-2020-jfm, villermaux-2022-pnasnexus, deike-2022-arfm}. For a single droplet, further increasing $\mathrm{We}$ triggers hydrodynamic instabilities of short wavelengths, such as Kelvin-Helmholtz instabilities and Rayleigh-Taylor waves, leading to breakup regimes referred to as ``sheet thinning" and ``catastrophic" \citep{guildenbecher-2009-expfluids, jackiw-2021-jfm}. However, such breakup modes were not observed in the cough machine experiments. 

Our previous study \citep{kant-2023-prf} on droplet generation in a cough machine focused on the effects of fluid viscosity, where it was observed that with increasing viscosity, the droplet size distribution shifts towards smaller sizes. However, the mucosalivary fluid lining of respiratory tracts is much more complex than a simple Newtonian liquid; it consists of water, salt, proteins (e.g. mucin), surfactants, and potentially pathogens \citep{vejerano-2018-jrsi, bastola-2021-expertopindrugdeliv}. The inclusion of salt mainly affects the evaporation rate of the droplets \citep{seyfert-2022-prf}, whereas mucin imparts non-Newtonian viscoelastic properties to the liquid \citep{anwarulhasan-2010-jnnfm, hamed-2020-sm, rodriguezhakim-2022-sm}. These viscoelastic effects are known to hinder fragmentation in liquids by damping small-scale surface perturbations at the liquid-air interface \citep{hoyt-1974-jfm, haessig-2025-arxiv}, and promote the formation of ligaments during the final stage of shear-induced breakup \citep{wilcox-1961-japplpolymsci, chandra-2023-jfm}. In contrast, in the ligament-mediated breakup regime, viscoelastic liquids can generate larger droplets by accumulating larger volumes in the `beads' of the characteristic `beads-on-a-string' structures prior to breakup \citep{clasen-2006-jfm, bhat-2010-natphys, keshavarz-2016-prl, sen-2022-arxiv, chandra-2024-prf}. However, the existing literature on the aerodynamic breakup of viscoelastic droplets have almost exclusively focused on the sheet thinning and catastrophic modes of breakup observed at high $\mathrm{We}$. Despite its obvious relevance in respiratory aerosol generation, little is known about the breakup modes or droplet size distribution of mucosalivary fluid-like viscoelastic liquids in a cough machine. 

Here, we investigate the effects of viscoelasticity on aerosol generation in a cough machine. Our experimental system resembles the one from our previous work \citep{kant-2023-prf} -- a thin liquid film is subjected to a strong interfacial shear by a high-speed air flow, resulting in the breakup of the sheet into droplets. In particular, we aim to address how the viscoelasticity of the mucosalivary fluid affects the droplet generation mechanism, and the size distribution of the droplets thus produced. This is experimentally achieved by dissolving a small amount of a long chain polymer in our aqueous mucosalivary-mimetic fluid. By changing the concentration of the dissolved polymers, we experimentally demonstrate how viscoelasticity shifts the droplet size distribution of the generated aerosol to smaller sizes. Further, using numerical simulations, we elucidate the underlying physical mechanisms dictating this shift in the size distribution. 

\section{Experimental methods} \label{sec:exp-methods}

\subsection{Experimental protocol} \label{subsec:exp-protocol}

\begin{figure}
    \centering
    \includegraphics[width=\textwidth]{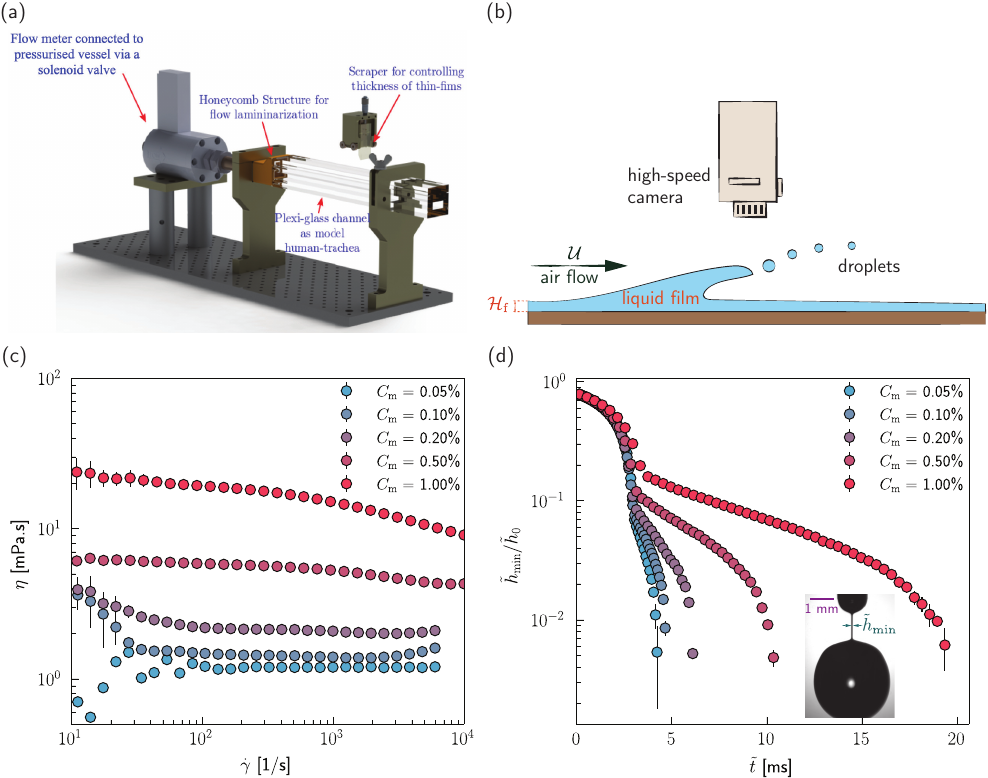}
    \caption{\textbf{Experimental setup and fluid characterization} \textbf{(a)} Schematic of the laboratory-scale cough machine, reproduced from our previous work \citep{kant-2023-prf}. \textbf{(b)} Schematic depiction of the experimental configuration. \textbf{(c)} Measured variation of the dynamic viscosity $\eta$ of mucosalivary-mimetic fluids, of different polymer concentrations $C_{\mathrm{m}}$, with shear rate $\dot{\gamma}$. \textbf{(d)} Temporal ($\tilde{t}$) variation of the minimum filament thickness $\tilde{h}_{\min}$, normalized by the initial filament thickness $\tilde{h}_{0}$, for different polymer concentrations $C_{\mathrm{m}}$; the inset depicts a typical experimental snapshot of a stretching filament (for $C_{\mathrm{m}}$ = 0.50\%), where the minimum filament thickness $\tilde{h}_{\min}$ is marked. In each panel, the discrete data markers denote the mean of three independent experimental realizations while the error bars indicate one standard deviation.}
    \label{fig:fig-1}
\end{figure}

The experimental setup (as shown in figure \ref{fig:fig-1}a), comprising of the laboratory-scale cough machine, is identical to the one used in our previous study \citep{kant-2023-prf}. Briefly, the experimental configuration consists of a 30 cm long model trachea made of transparent plexi-glass, with a 2 cm wide and 1 cm high rectangular cross-section. The base of this rectangular channel is covered by a thin layer of mucosalivary-mimetic fluid. A sharp-edged scraper running over a rail-guide is used to create a uniform thin film of the fluid with the desired thickness. The upstream of the channel is connected to a pressurized tank, where a valve is actuated to create an airflow mimicking a typical cough, whereas the other end of the channel is open to atmosphere. While the valve is open (for a duration of 400 ms), the air flow rate ramps-up steeply to its maximum, and then drops down -- mimicking a real cough \citep{king-1985-japplphysiol}. A honeycomb structure at the channel inlet homogenizes the incoming flow. The free-stream velocity $\mathcal{U}$ (= 30 m/s, kept unchanged throughout the experiments) was measured at the inlet of the channel via a flow meter (EL-Flow Select, Bronkhorst). 

The atomization of the liquid thin film inside the channel was recorded via a high-speed camera (Nova S12, Photron) at a maximum frame rate of 40000 frames-per-second (schematically depicted in figure \ref{fig:fig-1}b). Size statistics of the droplets resulting from atomization was measured using a two-component Phase Doppler Anemometry (PDA) system with a 112~mm FiberFlow probe (Dual PDA, Dantec Dynamics). Details of the measurement principle have already been described elsewhere \citep{book-albrecht, qiu-1992-expfluids}. The measurement volume for PDA was fixed at $\approx$ 15 cm downstream of the channel exit along its centerline. Note that this choice of measurement location inherently introduces a bias towards the smaller droplet sizes in the measured distributions, since the larger droplets are likely to ballistically fall out of the expelled turbulent puff \citep{wells-1936-jama, chong-2021-prl, ng-2021-prf}.

\subsection{Characterization of the mucosalivary-mimetic fluid} \label{subsec:fluid-prop}

The mucosalivary-mimetic fluids used in the present study closely mimic the viscoelastic rheology of real mucosalivary fluids \citep{zussman-2007-jdentres, lai-2009-advdrugdelivrev, haward-2011-rheolacta, wagner-2017-jrheol, hamed-2020-sm}. Aqueous solutions of polyethylene oxide (average molecular weight~$\approx$~6~$\times$~10$^{\text{5}}$~Da, Sigma-Aldrich, henceforth referred to as ``PEO"), of concentrations $C_{\mathrm{m}}$ (by mass) ranging from 0.05\% to 1\% were prepared by adding the required amount of polymer powder to purified water (Milli-Q), and stirring with a magnetic stirrer for 24 hours. The small amount of polymeric addition did not significantly alter the liquid density, and all solutions had the same density of $\rho_{\mathrm{l}}=$ 993 kg/m$^3$. The liquid-air interfacial tension coefficient (surface tension) was measured using the pendent drop method on a commercial drop shape analyzer (DSA 100E, Kr\"{u}ss GmBH) \cite{berry2015measurement, stauffer1965measurement, hansen1991surface}. All polymeric solutions had similar values of $\gamma = 62$ mN/m, whereas the $C_{\mathrm{m}}=0\%$ case had a higher $\gamma = 72$ mN/m. 

\begin{table}
	\begin{tabular}{ccc}
		\hline
		$C_{\mathrm{m}}$ [wt.\%] & $\lambda$ [ms] & $\eta_{\mathrm{N}}$ [mPa.s] \\
		\hline
		0 & - & 0.9 \\
		0.05 & 0.26 & 1.2 \\
		0.10 & 0.41 & 1.4 \\
		0.20 & 0.62 & 2.1 \\
		0.50 & 1.36 & 5.8 \\
		1.00 & 2.58 & 18.0 \\
		\hline
	\end{tabular}
	\caption{Measured relaxation time $\lambda$ and plateau value of dynamic viscosity $\eta_{\mathrm{N}}$ for polymeric liquids of concentration $C_{\mathrm{m}}$.} 
	\label{tab:tab-1}
\end{table}

A rotational stress-controlled rheometer (MCR 502, Anton Paar GmbH) with a cone-and-plate configuration (\SI{1}{\text{\degree}}, 50 mm diameter, and mean gap of 0.1 mm) was used to measure the shear viscosities of the samples. The detailed shear viscosity vs. shear rate curves are shown in figure \ref{fig:fig-1}c. The plateau viscosity value, $\eta_{\mathrm{N}}$, which denotes the viscosity that remains practically invariant with shear rate $\dot{\gamma}$, is summarized in table \ref{tab:tab-1}. All measurements were performed at 19\SI{}{\text{\celsius}} and in triplicate. The Ohnesorge number based on the plateau viscosity $\eta_N$ and initial film thickness $\mathcal{H}_{\mathrm{f}}$ is defined as $\mathrm{Oh} = \eta_N/\left(\rho_{\mathrm{l}} \gamma \mathcal{H}_{\mathrm{f}}\right)^{1/2}$. The Ohnesorge number range achieved in the present study is $\mathrm{Oh} = 0.003 - 0.07$, which is on the lower end of the range covered by \citet{kant-2023-prf}. The focus of the present work is on elucidating the role of the viscoelasticity on the atomization phenomena observed during a coughing or sneezing event. Most of the fluids ($C_{\mathrm{m}} <$ 0.50\%) used in our experiments behaved as Boger fluids \citep{james-2009-arfm}, i.e. their shear viscosity remained independent of the shear rate (see figure \ref{fig:fig-1}c). However, dissolved polymers may also impart shear thinning behavior, as observed for $C_{\mathrm{m}}$ = 0.50\% and 1.00\% (see figure \ref{fig:fig-1}c), which further complicates an already complex phenomenon.

The relaxation times, $\lambda$, of the polymeric liquids were measured from the extensional thinning of liquid filaments in a pendent droplet configuration \citep{deblais-2018-prl, deblais-2020-jfm, eggers-2020-jfm, sen-2021-jfm, sen-2022-arxiv}. The pendent droplets were generated by connecting a stainless steel blunt tip dispensing needle (inner diameter = 0.84 mm, Nordson EFD) to a syringe (Inkjekt 5 mL, Braun) via a flexible plastic PEEK tubing (Upchurch Scientific). The syringe was actuated by operating a syringe pump (Harvard Apparatus) at a very low flow rate. The temporal evolution of the thinning filament was visualized from the side by a high-speed camera (at 10000 frames-per-second, SA-X2, Photron) connected to a telecentric lens (Navitar), while the filament was back-illuminated by a cold LED light source (KL 2500, Schott). A typical experimental snapshot of the stretching filament is shown in the inset of figure \ref{fig:fig-1}d. When the elastic and capillary forces balance each other, the minimum thickness of the filament, $\tilde{h}_{\min}$, decays exponentially \citep{deblais-2018-prl, deblais-2020-jfm, bousfield-1986-jnnfm, anna-2001-jrheol, eggers-2020-jfm, sen-2022-arxiv} with time $\tilde{t}$: $\tilde{h}_{\min} \sim \exp \left( - \tilde{t} / 3 \lambda \right)$, as also observed in figure \ref{fig:fig-1}d. Thus, the relaxation time $\lambda$ can be calculated from the experimentally-measured thinning dynamic. These measurements have been reported in table \ref{tab:tab-1}. All measurements were performed at 19\SI{}{\text{\celsius}} and in triplicate. 

\section{Experimental results} \label{sec:exp-results}

\subsection{Fragmentation of polymeric liquid sheets occurs via bag formation} \label{subsec:fragmentation}

\begin{figure}[h]
    \centering
    \includegraphics[width=\textwidth]{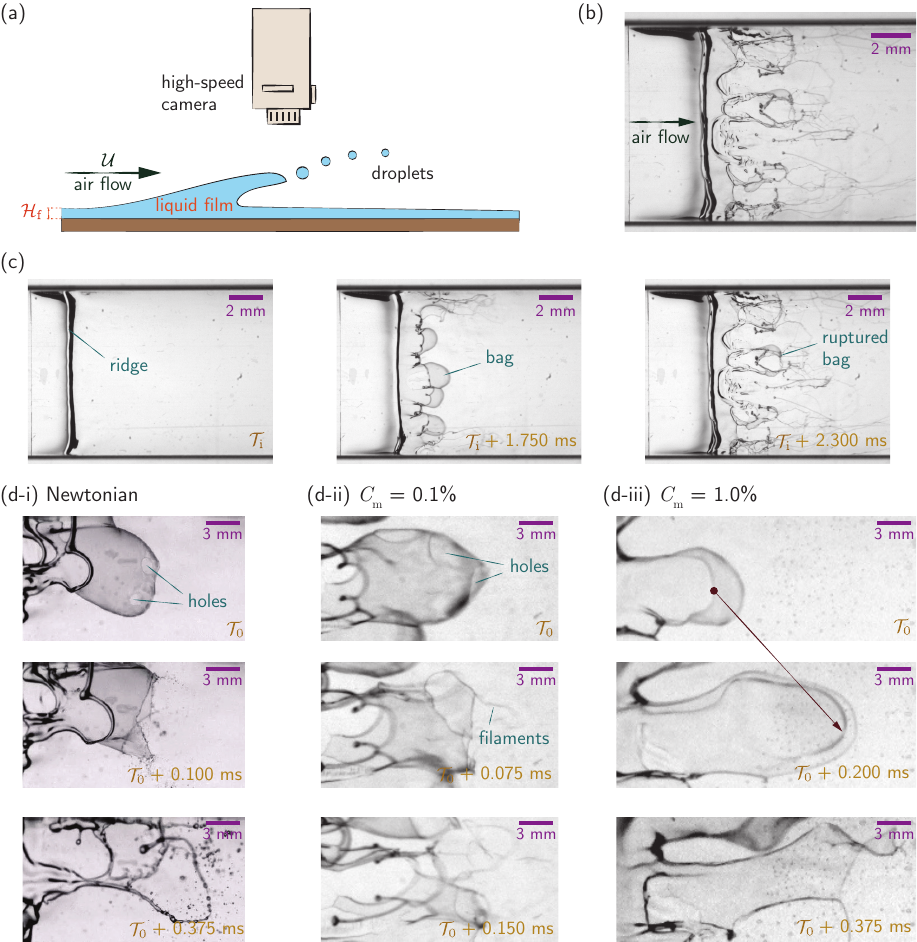}
    \caption{\textbf{Fragmentation of a sheared liquid film in a cough machine.} \textbf{(a)} Time-lapsed experimental snapshots of the atomization of a sheared liquid sheet in a cough machine for concentration $C_{\mathrm{m}}$ = 0.1\%. Time-lapsed experimental snapshots of the bag breakup observed in a \textbf{(b-i)} Newtonian liquid sheet,  and polymeric liquid sheets of concentrations \textbf{(b-ii)}~$C_{\mathrm{m}}$~=~0.1\% and \textbf{(b-iii)} $C_{\mathrm{m}}$ = 1.0\%, where the air flow direction is from left to right. The arrow in panel d-iii highlights the shear-induced stretching and flattening of the liquid sheet enclosing a bag. The Newtonian observations in panel b-i, corresponding to a water-glycerol mixture of viscosity 5 mPa.s, has been adopted from our previous work \citep{kant-2023-prf}. See movie S1 in the supplementary information for the corresponding movie.}
    \label{fig:fig-2}
\end{figure}

A liquid film of uniform initial thickness $\mathcal{H}_{\mathrm{f}}$ = 1 mm, having a concentration $C_{\mathrm{m}}$ (by mass) of dissolved polymers (PEO), is deposited onto the bottom of the cough machine channel. A high-velocity air stream, flowing with a free-stream velocity $\mathcal{U}$ = 30 m/s over the top of the liquid film, causes the liquid film to fragment into droplets, which is imaged from the top with a high-speed camera (as shown in figure \ref{fig:fig-1}b). A sequential overview of the fragmentation of the liquid film is shown in figure~\ref{fig:fig-2}a, which is described in detail in our previous study \citep{kant-2023-prf} with Newtonian liquids. Briefly, the high-velocity air stream results in a shear-induced Kelvin-Helmholtz instability of the liquid film \citep{scardovelli-1999-arfm, hoepffner-2011-prl}, characterized by creases, leading to the formation of a ridge at time $\mathcal{T}_{\mathrm{i}}$. This ridge subsequently inflates into several liquid bags via a Rayleigh-Taylor-like mechanism \citep{kant-2023-prf}. These bags, while remaining attached at their downstream end to the liquid film, continue to inflate under aerodynamic forces, and eventually rupture and atomize into small droplets that are expelled out of the channel. A typical fragmentation of a liquid bag happens over a time scale of~$\mathcal{O}$(1~ms) in the present experiments. 

The fragmentation of a single liquid bag for Newtonian and viscoelastic liquids is examined more closely in the time-lapsed sequence of experimental snapshots shown in figures \ref{fig:fig-2}b-i -- \ref{fig:fig-2}b-iii. We start with examining the classical Newtonian case (corresponding to a water-glycerol mixture of viscosity 5 mPa.s \citep{kant-2023-prf}), as shown in figure~\ref{fig:fig-2}b-i. As the inflating liquid bags destabilize, holes nucleate on the bags, primarily at the thinnest parts of the liquid film encompassing the bag \citep{villermaux-2020-jfm, lohse-2020-pnas}. While the exact physical mechanisms dictating the initial nucleation of these holes are still a topic of intense scientific debate \citep{taylor-1973-jfm, villermaux-2007-arfm, villermaux-2020-jfm, lohse-2020-pnas}, and may vary from case to case, we speculate that here the pressure gradient across the liquid film in the present experiments might at least partially be responsible for local heterogeneities in the film thickness \citep{villermaux-2020-jfm, vledouts-2016-jfm, klein-2020-jfm, bremond-2005-jfm, kant-2023-prf}, leading to the nucleation of holes \citep{bazzi-2019-jnnfm}. These holes then expand at the classical Taylor-Culick retraction velocity \citep{villermaux-2009-natphys, savva-2009-jfm, lhuissier-2012-jfm, sanjay-2022-jfm, kant-2023-prf} (determined by the competition between the inertia of the liquid sheet and its surface tension), and the unstable motion of the retracting liquid sheet generates small droplets with diameters less than 50 \SI{}{\text{\micro} \text{m}}. The original rim of the bag, which continuously accumulates liquid as the hole expands, eventually breaks up to generate droplets with diameters greater than 200 \SI{}{\text{\micro} \text{m}}. These mechanisms for Newtonian liquids have already been observed, and discussed in detail, in our previous work \citep{kant-2023-prf}. 

When a small amount of polymers are present in the liquid ($C_{\mathrm{m}}$ = 0.1\%, figure \ref{fig:fig-2}b-ii), inflating bag morphologies similar to the Newtonian case also appear, upon which again holes nucleate. However, when these bags rupture, thin liquid filaments, instead of small droplets, emanate from the expanding holes (figure \ref{fig:fig-2}b-ii). Similar filamentous structures during the fragmentation of viscoelastic liquids have been observed in other studies~\citep{park-2008-atsprays, christanti-2006-atsprays, thompson-2007-jnnfm, karim-2018-jnnfm, gaillard-2022-jnnfm} as well. Surprisingly, when the amount of dissolved polymers is increased even further ($C_{\mathrm{m}}$ = 1.0\%, figure \ref{fig:fig-2}b-iii), we observe an entirely new mode of breakup, which is absent in the previously studied Newtonian cases \citep{kant-2023-prf}: some bags break down when the top sheet flips to the downstream direction, which is accompanied by a shape transition from a liquid bag to a flattened liquid sheet (as marked by the arrow in figure \ref{fig:fig-2}b-iii). The rims of this sheet then retract and merge into a thicker ligament, which eventually falls back into the initial liquid film. This highly complex fragmentation behavior could be due to the liquid sheet's increased resistance to the nucleation of weak spots or holes, resulting in the incoming air flow escaping around the rim of the bag. However, the higher polymer concentration ($C_{\mathrm{m}}$ = 1.0\%) in the case depicted in figure \ref{fig:fig-2}b-iii not only imparts viscoelasticity to the liquid, but also a shear thinning rheology (as evident from figure \ref{fig:fig-1}c). This shear thinning rheology may also play a significant role at the characteristic high shear rates of the present experiments, but quantifying the precise nature of this role is beyond the scope of the present work. 

\subsection{Viscoelasticity results in highly stretched liquid bags} \label{subsec:bag-dynamics}

\begin{figure}
    \centering
    \includegraphics[width=\textwidth]{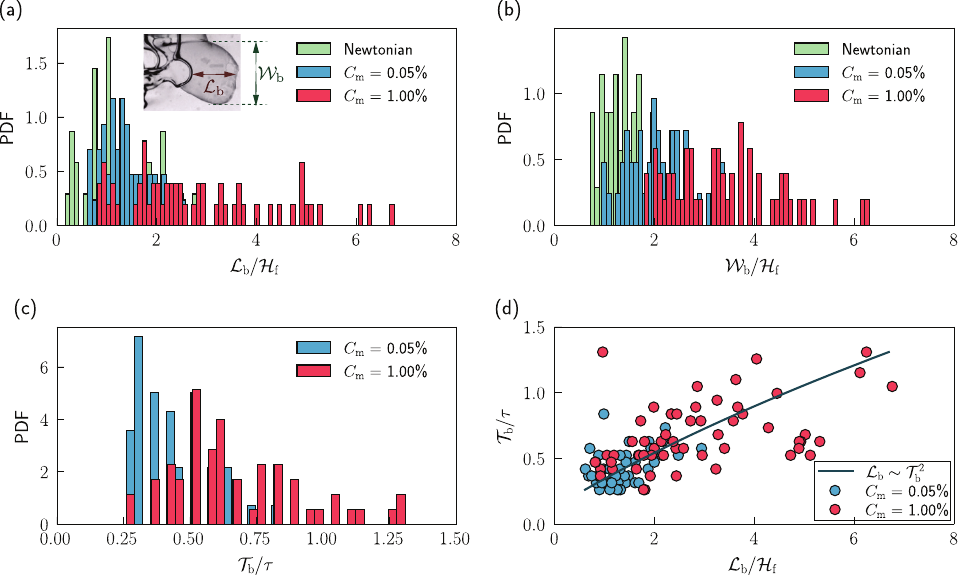}
    \caption{\textbf{Size and lifetime of bags during fragmentation.} Distribution of \textbf{(a)} lengths ($\mathcal{L}_{\mathrm{b}}$) and \textbf{(b)} widths ($\mathcal{W}_{\mathrm{b}}$) of bags, normalized by the initial liquid film thickness $\mathcal{H}_{\mathrm{f}}$, just prior to breakup; the inset in panel a depicts a typical measurement of $\mathcal{L}_{\mathrm{b}}$ and $\mathcal{W}_{\mathrm{b}}$. The Newtonian data in panels a and b, corresponding to a water-glycerol mixture of viscosity 5 mPa.s, have been adopted from our previous work \citep{kant-2023-prf}. \textbf{(c)} Distribution of the lifetimes ($\mathcal{T}_{\mathrm{b}}$) of bags, normalized by the characteristic time scale \citep{villermaux-2009-natphys} $\tau_{\mathrm{f}} = \mathcal{H}_{\mathrm{f}} \sqrt{\Gamma_{\rho}} / \mathcal{U}$. \textbf{(d)} Scatter map of the normalized lengths and lifetimes of bags, where the solid line denotes $\mathcal{L}_{\mathrm{b}} \sim \mathcal{T}_{\mathrm{b}}^{2}$ \citep{villermaux-2009-natphys}.}
    \label{fig:fig-3}
\end{figure}

While the shear-induced fragmentation of a polymeric sheet still occurs via an intermediate bag formation (for dilute polymer concentrations, as shown in figure \ref{fig:fig-2}), the dynamics of bag deformation and breakup can be markedly different. We quantify this by measuring the length $\mathcal{L}_{\mathrm{b}}$ and width $\mathcal{W}_{b}$ of bags just prior to breakup; a typical measurement is shown in the inset of figure \ref{fig:fig-3}a. Our previous study \citep{kant-2023-prf} demonstrated that both the length and width of liquid bags increase with increasing viscosity, i.e. bags of highly viscous liquids can be inflated to a greater extent prior to fragmentation. The distributions of the lengths and widths of the polymeric liquid bags in the present work, normalized by the initial liquid film thickness $\mathcal{H}_{\mathrm{f}}$, are shown in figures \ref{fig:fig-3}a and \ref{fig:fig-3}b, respectively, for two different polymer concentrations ($C_{\mathrm{m}}$ = 0.05\% and 1.00\%). The corresponding size distributions for a Newtonian liquid (water-glycerol mixture of viscosity 5 mPa.s), as observed in our previous work \citep{kant-2023-prf}, are also shown in figures \ref{fig:fig-3}a and \ref{fig:fig-3}b. Note that while the $C_{\mathrm{m}}$ = 0.05\% liquid has a lower viscosity (1.2~mPa.s, see figure \ref{fig:fig-1}c and table \ref{tab:tab-1}) than the Newtonian liquid (5 mPa.s), it exhibits larger lengths and widths of the liquid bags. This observation is in contrast to the expectation from our previous work \citep{kant-2023-prf}: a lower viscosity (i.e. lower Oh) should have resulted in smaller bags. Hence, figures \ref{fig:fig-3}a and \ref{fig:fig-3}b demonstrate that the viscoelasticity imparted by the presence of even a small amount of dissolved polymers (concentration $C_{\mathrm{m}}$ = 0.05\% in this case) can have a stabilizing effect on the liquid bags, causing the bags to be inflated to a greater extent before they eventually fragment. Further increasing the polymer concentration can result in even larger bags, with some bags, remarkably, reaching lengths and widths more than 6 mm (for $C_{\mathrm{m}}$ = 1.00\%). However, and as mentioned before, this dramatic effect at high concentrations cannot only be attributed to the stronger viscoelastic response of the $C_{\mathrm{m}}$ = 1.00\% liquid as compared to the $C_{\mathrm{m}}$~=~0.05\% liquid (see figure \ref{fig:fig-1} and table \ref{tab:tab-1}). The $C_{\mathrm{m}}$ = 1.00\% liquid exhibits both a higher zero-shear viscosity as well as a shear thinning rheology (see figure \ref{fig:fig-1}c and table \ref{tab:tab-1}), and the effect of the latter on the dynamics of bag breakup are yet to be fully understood. 

The lifetime, $\mathcal{T}_{\mathrm{b}}$, of a liquid bag is defined as the time duration between the formation of the crease on the sheared liquid sheet to the appearance of the first hole on the surface of the bag. The measured lifetime in the present experiments is normalized by the characteristic timescale \citep{villermaux-2009-natphys} $\tau_{\mathrm{f}} = \mathcal{H}_{\mathrm{f}} \sqrt{\Gamma_{\rho}} / \mathcal{U}$, where $\Gamma_{\rho} = \rho_{\mathrm{l}} / \rho_{\mathrm{g}}$ denotes the ratio of densities ($\rho$) of liquid (subscript ``$\mathrm{l}$") and air (subscript ``$\mathrm{g}$"), and its distribution is shown in figure \ref{fig:fig-3}c for two different polymeric concentrations. Bags with a higher concentration exhibit longer lifetimes and reach a larger size before fragmentation. The latter is also evident from the scatter map shown in figure~\ref{fig:fig-3}d, where the lifetime and lengths appear to be positively correlated. Considering that the inflation of a bag under aerodynamic loads follows \citep{villermaux-2009-natphys} $\mathcal{L} \sim \mathcal{T}^{2}$, where $\mathcal{L}$ and $\mathcal{T}$ denote length and time, respectively, a longer lifetime would indeed allow a bag to inflate to a larger size. 

\subsection{Polymers reduce the size of the expelled droplets} \label{subsec:size-distribution}

\begin{figure}
    \centering
    \includegraphics[width=\textwidth]{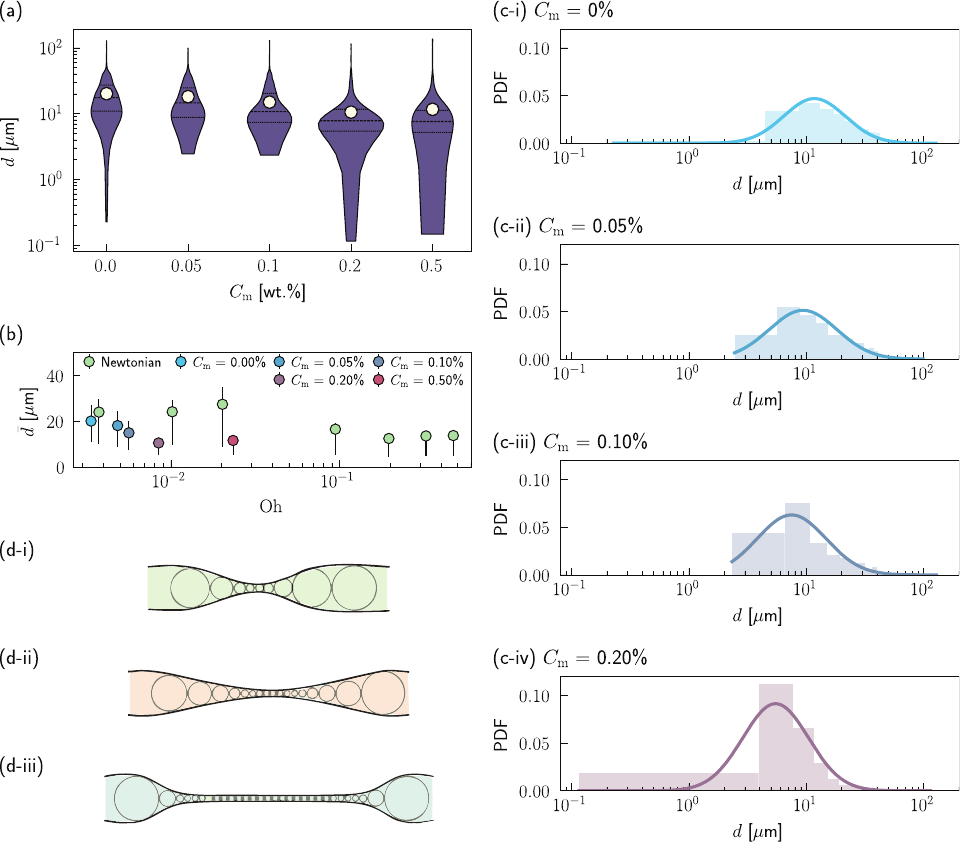}
    \caption{\textbf{Size distributions of the expelled droplets from the cough machine.} \textbf{(a)} Distributions of the droplet diameter, $d$, at the channel exit for different polymeric concentrations, $C_{\mathrm{m}}$, where the discrete data markers denote the mean value of each distribution. \textbf{(b)} Variation of mean droplet diameter $\bar{d}$ with the Ohnesorge number $\mathrm{Oh}$, where the ``Newtonian" data corresponds to the data presented in figure 9 of \citet{kant-2023-prf} and the error bars are at 20- and 80-percentile of $d$. Distributions of the droplet diameter at the channel exit for \textbf{(c-i)}~$C_{\mathrm{m}}$~=~0\% (pure water), \textbf{(c-ii)} $C_{\mathrm{m}}$ = 0.05\%, \textbf{(c-iii)} $C_{\mathrm{m}}$ = 0.10\%, and \textbf{(c-iv)} $C_{\mathrm{m}}$ = 0.20\%, where the histograms denote the experimental measurements and the solid lines indicate the corresponding fits to the log-normal distribution \citep{kant-2023-prf}. Schematics of \textbf{(d-i)} Newtonian (of viscosity $\eta_{1}$), \textbf{(d-ii)} Newtonian (of viscosity~$\eta_{2}~>~\eta_{1}$), and \textbf{(d-iii)} viscoelastic liquid filaments prior to breakup, where the circles denote the protoblobs fitting in the filament thickness profile.}
    \label{fig:fig-4}
\end{figure}

The dynamics of the bag stretching, and the final dimensions of the bag prior to breakup, are known to have a strong bearing on the size distribution of the expelled droplets \citep{jackiw-2022-jfm, kant-2023-prf}. Thus far, we have demonstrated that viscoelasticity due to the presence of dissolved polymers can significantly increase the lifetime and, concomitantly, the size of the bags prior to atomization (see figure \ref{fig:fig-3}). These bags eventually fragment into droplets, and the size (diameter) distribution of these expelled droplets, measured via Phase Doppler Anemometry (PDA), is shown in figure \ref{fig:fig-4}. With increasing polymer concentration $C_{\mathrm{m}}$, the mean droplet diameter initially decreases before a slight increase again for $C_{\mathrm{m}}$ = 0.50\%, as shown in figure \ref{fig:fig-4}a.

We further compare the mean droplet diameter $\bar{d}$ in the present work with the results of \citet{kant-2023-prf} in figure \ref{fig:fig-4}b, where the Ohnesorge number is defined as $\mathrm{Oh} = \eta_N/\left(\rho_{\mathrm{l}} \gamma \mathcal{H}_{\mathrm{f}}\right)^{1/2}$. The mean diameter of the present $C_{\mathrm{m}}$ = 0\% case compares favorably with the least viscous (lowest Oh) case in \citet{kant-2023-prf}. With increasing $C_{\mathrm{m}}$, $\bar{d}$ reduces, and the resulting $\bar{d}$ is much smaller than that of the Newtonian fluids at similar $\mathrm{Oh}$. Remarkably, cases $C_{\mathrm{m}}$ = 0.10\% and 0.20\% produce droplets as small ($\bar{d} \approx 10$~\SI{}{\text{\micro} \text{m}}) as those from Newtonian fluids with an $\mathrm{Oh}$ more than 10 times larger.

The non-monotonicity for $C_{\mathrm{m}}$~=~0.50\% may be attributed to the shear thinning behavior observed at high shear rates for $C_{\mathrm{m}}$~=~0.50\% (see figure \ref{fig:fig-1}c). The droplet size distribution for the Newtonian ($C_{\mathrm{m}}$ = 0\%) and the Boger polymeric fluids (viscosity independent of shear rate \citep{james-2009-arfm}, see figure \ref{fig:fig-1}c) are shown in figures \ref{fig:fig-4}c-i -- \ref{fig:fig-4}c-iv, where the histograms denote the experimental measurements and the solid lines indicate the corresponding fits to the log-normal distribution \citep{kant-2023-prf}. For the Newtonian case ($C_{\mathrm{m}}$ = 0\%, figure \ref{fig:fig-4}c-i), the droplet size distribution peaks around 20~\SI{}{\text{\micro} \text{m}}, which is in agreement with our previous study \citep{kant-2023-prf}. However, as the polymer concentration $C_{\mathrm{m}}$ increases, the peak of the distribution gradually shift towards smaller values ($\lesssim$~10~\SI{}{\text{\micro} \text{m}}, figures \ref{fig:fig-4}c-ii -- \ref{fig:fig-4}c-iv). Note that no droplets were detected for $C_{\mathrm{m}}$ = 1.00\% as the liquid was expelled mostly in the form of filaments and not droplets. 

We note that most prior studies on the fragmentation of viscoelastic liquids report an \emph{increase} in the droplet mean diameter with increasing polymer concentration \citep{keshavarz-2015-jnnfm, christanti-2006-atsprays, gaillard-2022-jnnfm}, which is an apparent contradiction to our observations of a \emph{decrease} reported in figure \ref{fig:fig-4}. This discrepancy may be attributed to two factors. Firstly, the droplet population captured by the PDA primarily originates from the unsteady retraction of the bursting micrometric thin liquid film \citep{kant-2023-prf}, while the larger droplets (diameter in the order of a few hundred micrometres) generated by the breakup of filaments tend to fall out of the expelled turbulent puff \citep{wells-1936-jama, chong-2021-prl, ng-2021-prf}. In contrast, the entire population of droplets formed in the atomization process, or only the large droplets if measured via high-speed imaging, have been taken into consideration in the previous studies. Secondly, in the current study, the presence of viscoelasticity dramatically alters the liquid sheet (or filament) morphology prior to fragmentation, as illustrated schematically in figures~\ref{fig:fig-4}d-i~--~\ref{fig:fig-4}d-iii. The necking region of a Newtonian liquid filament exhibits a conical shape (figure~\ref{fig:fig-4}d-i), whose axial extent increases with increasing viscosity (figure \ref{fig:fig-4}d-ii) \citep{eggers-1993-prl, papageorgiou-1995-pof}. In contrast, for a viscoelastic liquid (figure \ref{fig:fig-4}d-iii), the filament thickness decreases exponentially with time \citep{deblais-2018-prl, deblais-2020-jfm, eggers-2020-jfm, sen-2022-arxiv}, which is much slower compared to the linear (viscous \citep{eggers-1993-prl, papageorgiou-1995-pof}) or 2/3-power law (potential flow \citep{ting-1990-siamjapplmath}) temporal decreases observed for Newtonian liquids. Moreover, the viscoelastic liquid exhibits little variation of filament thickness in the lateral direction, unlike a Newtonian filament. This near-uniform stretching of the filament delays (or even inhibits) pinch-off \citep{amarouchene-2001-prl, sen-2021-jfm}, allowing the filament to reach a greater length, thus leading to larger but thinner bags. When a thin liquid filament (or sheet) atomizes, the size distribution of the resulting droplets is largely determined by the local thickness of the liquid structures \citep{marmottant-2004-pof, keshavarz-2016-prl} (as marked by the circular protoblobs in figures \ref{fig:fig-4}d-i -- \ref{fig:fig-4}d-iii), and the mean diameters of the droplets produced by thinner sheets also tend to be smaller, which explains the shift towards smaller droplet sizes at higher viscoelasticity (figure \ref{fig:fig-4}a) and viscosity \citep{kant-2023-prf}. We give further support to this claim in the following section with the aid of numerical simulations. 

\section{Numerical simulations of viscoelastic bag breakup} \label{sec:numerics}
\begin{figure}
    \centering
    \includegraphics[width=0.5\textwidth]{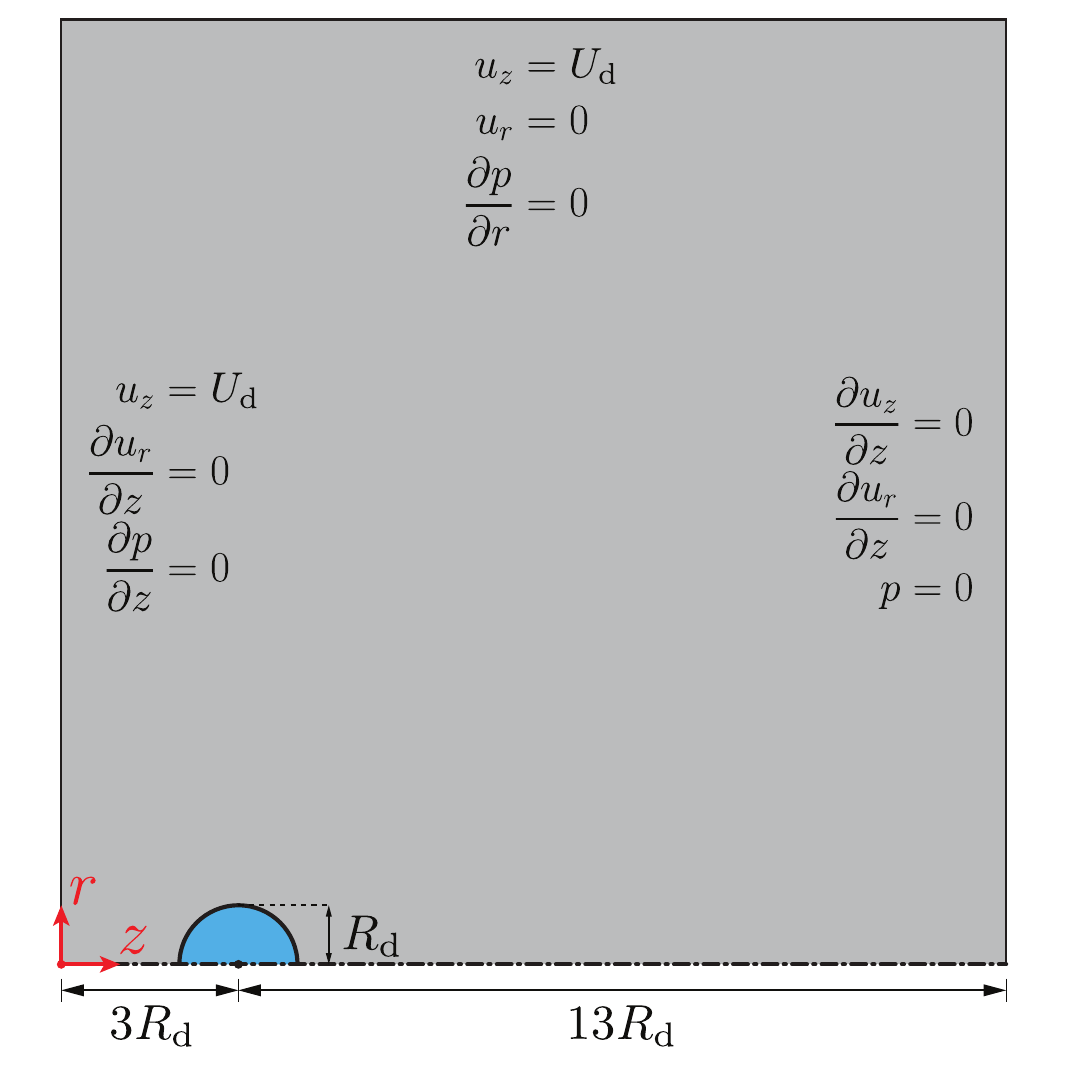}
    \caption{Schematic of the axisymmetric numerical domain (to scale), with the boundary conditions specified. The bottom boundary is the axis of symmetry, depicted by the dot-dashed line. The blue and gray colors depict the liquid and gas phases, respectively.}
    \label{fig:fig-5}
\end{figure}

\subsection{Dimensionless numbers, governing equations and numerical methods} \label{subsec:num-methods}

The breakup modes for the Newtonian and viscoelastic cases described thus far bear a strong resemblance to the bag breakup regime of aerodynamic atomization of a droplet \citep{villermaux-2009-natphys, reyssat-2007-epl, marcotte-2019-prf, jackiw-2021-jfm, hsiang-1992-ijmf, kulkarni-2014-pof, guildenbecher-2009-expfluids, chandra-2024-prf}. In the present work, the fragmentation morphology resembles the bag breakup mode, and multiple bags can form along the spanwise direction (as shown in figure \ref{fig:fig-2}a). The (film) Weber number $\mathrm{We}_{\mathrm{f}} = \rho_{\mathrm{g}} \mathcal{U}^{2} \mathcal{H}_{\mathrm{f}} / \gamma \approx$ 17, where $\gamma$ is the liquid-air interfacial tension coefficient. This Weber number falls within the range of $11\lesssim \mathrm{We}_{\mathrm{f}}\lesssim18$, which further confirms the bag breakup regime \citep{guildenbecher-2009-expfluids, hsiang-1995-ijmf}. Hence, to further elucidate the bag thinning and breakup process for viscoelastic liquids, we perform axisymmetric numerical simulations of a droplet in an impulsively-started air flow. In the absence of gravity, the dimensionless numbers that govern this flow are the Reynolds number $\mathrm{Re}_{\mathrm{d}} = \rho_{\mathrm{l}} U_{\mathrm{d}} R_{\mathrm{d}} / \eta_{\mathrm{l}}$, the (droplet) Weber number $\mathrm{We}_{\mathrm{d}} = \rho_{\mathrm{l}} U_{\mathrm{d}}^{2} R_{\mathrm{d}} / \gamma$, the density ratio~$\Gamma_{\rho} = \rho_{\mathrm{l}} / \rho_{\mathrm{g}}$, and the viscosity ratio $\Gamma_{\eta} = \eta_{\mathrm{l}} / \eta_{\mathrm{g}}$, where $U_{\mathrm{d}}$ is the velocity of the air stream flowing over the droplet of radius $R_{\mathrm{d}}$, and $\eta_{\mathrm{l}}$ and $\eta_{\mathrm{g}}$ are the liquid and gas viscosities, respectively. Additionally, times, lengths, and velocities are normalized by $\tau_{\mathrm{d}} = R_{\mathrm{d}} / U_{\mathrm{d}}$, $R_{\mathrm{d}}$, and $U_{\mathrm{d}}$, respectively. When viscoelastic effects are taken into account, two additional dimensionless numbers come into play: the elastoinertial number $\Pi = G / \rho_{\mathrm{l}} U_{\mathrm{d}}^{2}$ and the Weissenberg number $\mathrm{Wi} = \lambda / \tau_{\mathrm{d}}$, where $G$ and $\lambda$ are the elastic modulus and relaxation time of the polymeric liquid, respectively. In the Newtonian limit, setting $\Gamma_{\rho}$ = 1110, $\Gamma_{\eta}$ = 90.9, $\mathrm{Re}_{\mathrm{d}}$~=~1090$\Gamma_{\rho} / \Gamma_{\eta}$, and $\mathrm{We}_{\mathrm{d}}$ = 7.5$\Gamma_{\rho}$ leads to the formation of well-pronounced bags that eventually break up into smaller droplets, similar to the ones in our experiments \citep{marcotte-2019-prf}. 

\begin{figure}
	\centering
	\includegraphics[width=\textwidth]{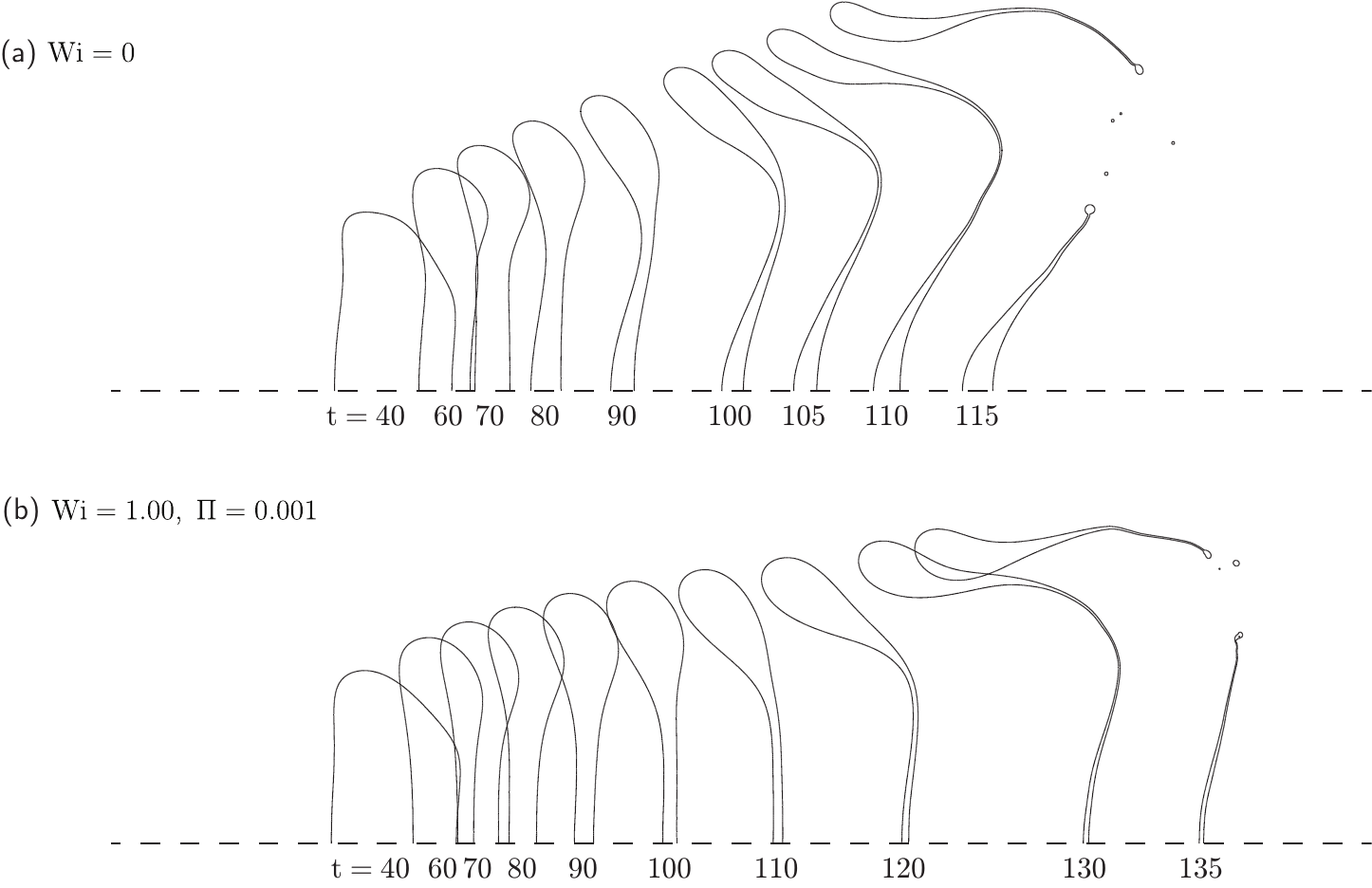}
	\caption{Sequence of interfacial shapes at different dimensionless times, $\mathrm{t}$, for \textbf{(a)} a Newtonian ($\mathrm{Wi}$ = 0) and \textbf{(b)} a viscoelastic droplet ($\mathrm{Wi}$~=~1.00, $\Pi$~=~0.001)}
	\label{fig:fig-6}
\end{figure}
\begin{figure}
	\centering
	\includegraphics[width=\textwidth]{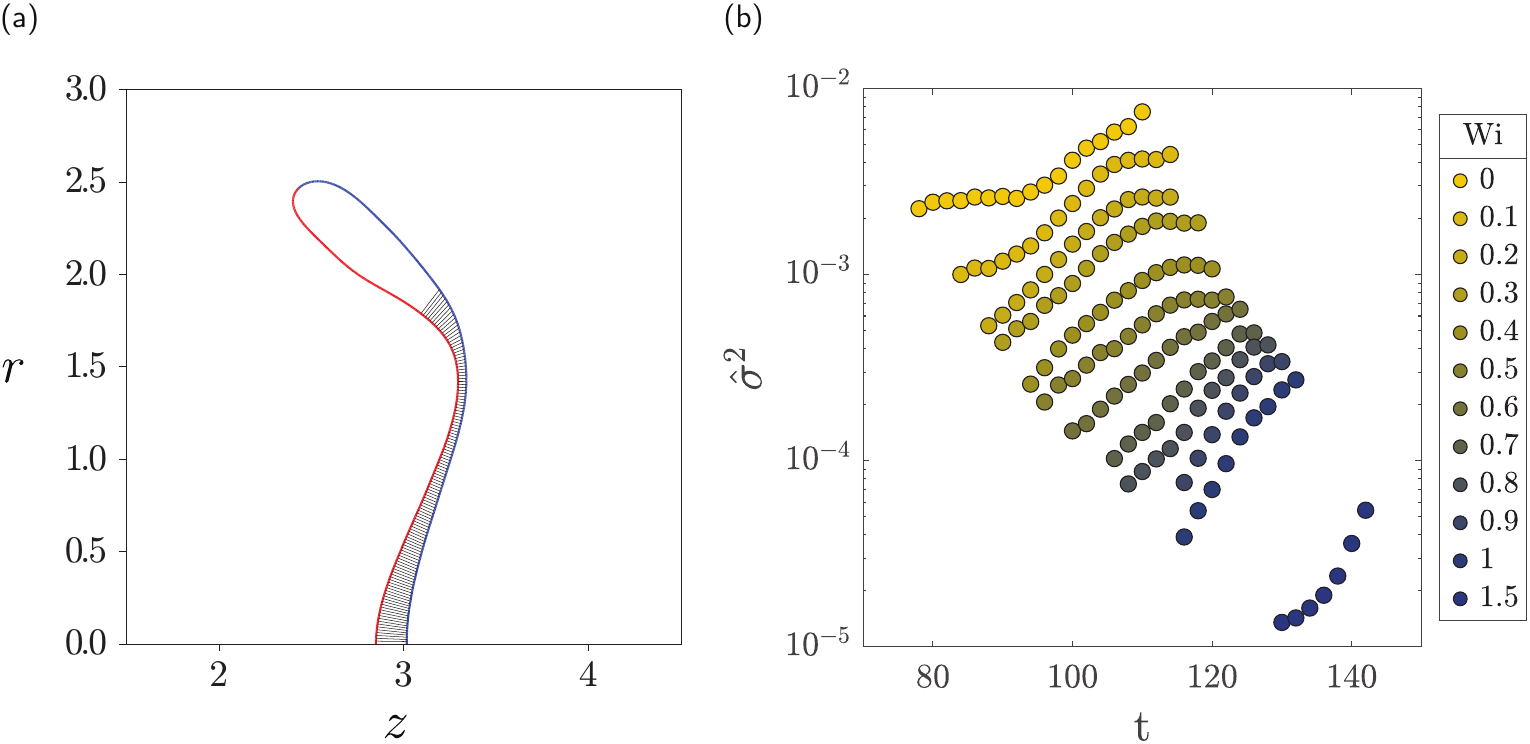}
	\caption{\textbf{(a)} Interface at $t$~=~100 for the Newtonian case ($\mathrm{Wi}$ = 0). This figure showcases the thickness extraction algorithm used for computing $\hat{\sigma}^2$. Equidistant points are sampled along the red part of the interface wherefrom normal lines emerge and intersect the opposite blue part. Segments above the minimum position, with a norm higher than the thickness at $r = 0$, are discarded. \textbf{(b)} Temporal evolution of the mean-squared deviation of the thickness profile of the interface about its minimum value, $\hat{\sigma}^{2}$ (see equation \eqref{eq:eq-variance}), at different $\mathrm{Wi}$ (see color code) for an elastoinertial number $\Pi$ = 0.001.}
	\label{fig:fig-7}
\end{figure}

The numerical setup and the boundary conditions are described in figure \ref{fig:fig-5}, where the air flow direction is from left to right. This reduced flow setup allows us to hone in on the thinning dynamics of the formed bags without considering the additional larger scale complexities of the flow inherent to the complete cough machine geometry. This also renders the simulations more accurate due to a better refinement.

The flow is fully described by the incompressible Navier-Stokes equations (conservation of mass and momentum), expressed in dimensionless form as
\begin{subequations}
	\begin{align}
		\nabla \cdot \mathbf{u} &= 0 \, , \label{eq:ns-a} \\
		\bar{\rho} \left( \frac{\partial \mathbf{u}}{\partial t} + \mathbf{u} \cdot \nabla \mathbf{u} \right) &= - \nabla p + \nabla \cdot \left( \bbsigma_{\mathrm{s}} + \bbsigma_{\mathrm{p}} \right) + \mathrm{We}^{-1} \kappa \delta_{s} \mathbf{n} \, , \label{eq:ns-b} 
	\end{align}
\end{subequations}
where $\bar{\rho}$ is the density, $\mathbf{u}$ the velocity field, and $p$ the pressure field. The viscous stress tensor $\bbsigma$ ($= \bbsigma_{\mathrm{s}} + \bbsigma_{\mathrm{p}}$) comprises of the Newtonian contribution of the solvent (subscript ``$\mathrm{s}$") and the viscoelastic contribution stemming from the addition of polymers (subscript ``$\mathrm{p}$"). The last term on the right-hand side of equation \eqref{eq:ns-b} represents the surface tension force, which only acts upon the interface. Hence, the Dirac $\delta_{s}$ function, defined as unity in the cells containing an interfacial segment and zero elsewhere, is invoked. Finally, $\kappa$ is the curvature of and $\mathbf{n}$ the normal to the interface. 

The viscous stress tensor pertaining to the solvent can be written as $\bbsigma_{\mathrm{s}} = 2 \bar{\eta} \mathbb{D}$, where $\bar{\eta}$ is the solvent viscosity and $\mathbb{D} = \left( \nabla \mathbf{u} + \left( \nabla \mathbf{u} \right)^{\mathrm{T}} \right) / 2$ the rate-of-strain tensor. The formulation of the polymeric component of the viscous stress tensor, $\bbsigma_{\mathrm{p}}$, requires choosing a constitutional rheological model relating the polymeric component of the stress to the strain rate. In the present work, we choose the Oldroyd-B model \citep{book-bird, oldroyd-1950-prsa, dixit-2024-arxiv}, due its proven capabilities in capturing the breakup of liquid sheets and filaments \citep{eggers-2020-jfm, sen-2021-jfm}. The Oldroyd-B model assumes a linear relationship between the elastic stresses and the polymeric deformations,
\begin{align}
	\bbsigma_{\mathrm{p}} &= \Pi \left( \mathbb{A} - \mathbb{I} \right) \, , 
	\label{eq:sigma-p}
\end{align}
where $\Pi$ denotes the previously-defined elastoinertial number, $\mathbb{A}$ the conformation tensor related to the deformation of polymeric chains, and $\mathbb{I}$ the identity tensor. The deformation and relaxation of the polymer chains are tracked in a temporal manner as
\begin{align}
	\stackrel{\nabla}{\mathbb{A}} &= - \frac{1}{\mathrm{Wi}} \left( \mathbb{A} - \mathbb{I} \right) \, ,
	\label{eq:oldroyd-b}
\end{align}
where
\begin{align}
	\stackrel{\nabla}{\mathbb{A}} &\equiv \frac{\partial \mathbb{A}}{\partial t} + \left(\mathbf{u} \cdot \nabla \right) \mathbb{A} - \mathbb{A} \cdot \nabla \mathbf{u} - \left( \nabla \mathbf{u} \right)^{\mathrm{T}} \cdot \mathbb{A} \, ,
	\label{eq:uc-ddt}
\end{align}
is the upper-convected time derivative of the conformation tensor $\mathbb{A}$. Equations \eqref{eq:sigma-p}, \eqref{eq:oldroyd-b}, and \eqref{eq:uc-ddt} can be combined to get an evolution equation of the polymeric stress tensor $\bbsigma_{\mathrm{p}}$, given by
\begin{align}
	\mathrm{Wi} \stackrel{\nabla}{\bbsigma}_{\mathrm{p}} + \bbsigma_{\mathrm{p}} &= 2 \mathrm{Re}_{\mathrm{p}}^{-1} \mathbb{D} \, ,
	\label{eq:uc-sigma-p}
\end{align}
where $\mathrm{Re}_{\mathrm{p}} = \left( \Pi \times \mathrm{Wi} \right)^{-1}$ is the polymeric Reynolds number. 

The equations described above are spatially discretized using a Finite Volume Method on a Cartesian grid, and then advanced in time. The nonlinear advection term in equation \eqref{eq:ns-b} is integrated using a momentum-conserving Bell-Colella-Glaz scheme \citep{bell-1989-jcp}, and the viscous contribution of the solvent is computed implicitly using a multigrid solver \citep{popinet-2015-jcp, saade-2023-jcp}. The log-conformation method \citep{fattal-2004-jnnfm, fattal-2005-jnnfm, lopezherrera-2019-jnnfm} is employed to circumvent the numerical instability arising at high Weissenberg numbers. Once calculated, the polymeric stress, $\bbsigma_{\mathrm{p}}$, is explicitly integrated and its contribution added to the momentum equation. The surface tension force is taken into account using a balanced-force continuum-surface-force (CSF) formulation \citep{brackbill-1992-jcp, popinet-2009-jcp}.

The interface, with a sharp representation, is tracked using a geometric Volume-of-Fluid (VoF) method \citep{scardovelli-1999-arfm}. The one-fluid formulation is adopted, where a volume fraction $c$ is used to delimit the two phases: $c$ is equal to 1 in purely liquid cells, to 0 in purely gaseous cells, and to intermediate values in mixture cells containing an interface. The volume fraction is advected with the local velocity field, hence obeying
\begin{align}
	\frac{\partial c}{\partial t} + \mathbf{u} \cdot \nabla c &= 0 \, . 
	\label{eq:vof-c}
\end{align}
Within this formulation, the fluid properties become a function of space, via the volume fraction, and are described as weighted arithmetic means: $\bar{\rho} = c + \Gamma_{\rho}^{-1} \left( 1 - c \right)$ and $\bar{\eta} = \mathrm{Re}^{-1} \left[ c + \Gamma_{\eta}^{-1} \left( 1 - c \right) \right]$. It must be noted that the polymeric properties are only defined in the liquid phase (weighted by $c$ in mixture cells), and taken as zero elsewhere.

The aforementioned methods are implemented in the free software program {\fontfamily{qzc}\selectfont Basilisk}~\citep{popinet-2015-jcp}, which is then used for the solution of the governing equations. {\fontfamily{qzc}\selectfont Basilisk} has quadtree discretization that allows for Adaptive Mesh Refinement (AMR) \citep{popinet-2015-jcp, popinet-2003-jcp} in the regions of interest, and thus for the accurate prediction of the thinning dynamics of the liquid bags.

\subsection{Stretched bags are thinner and lead to smaller droplets} \label{subsec:num-results}

Time-lapsed numerical snapshots of the deformation of a droplet in an impinging air flow, at different dimensionless time $t$-instants, are shown in figures \ref{fig:fig-6}a (for a Newtonian droplet) and \ref{fig:fig-6}b (for a viscoelastic droplet). The droplet starts thinning in the axial direction, while elongating in the radial direction due to mass conservation. Eventually, a bag-like structure is formed, which continues thinning, and culminates in a break up. One must be reminded, however, that the present numerical simulations are axisymmetric and therefore do not account for the out-of-plane instabilities that usually cause break up \citep{squire-1953-brjapplphys, bremond-2007-jfm}. Instead, the fragmentation herein occurs when the minimum thickness of the liquid filament becomes of the order of the local mesh size \citep{marcotte-2019-prf}. For the Newtonian scenario (figure \ref{fig:fig-6}a), at times $t >$ 100, the filament morphology is accurately described by the schematic representations in figures \ref{fig:fig-4}d-i and \ref{fig:fig-4}d-ii. The viscoelastic scenario shown in figure \ref{fig:fig-6}b corresponds to $\mathrm{Wi}$ = 1, which indicates that the relaxation time~$\lambda$ of the dissolved polymers is of the same order of the inertial time scale $\tau_{\mathrm{d}}$. However, even for $\mathrm{Wi}$ = 1, the precise value of~$\Pi$~(=~0.001 in figure \ref{fig:fig-6}b) has a strong bearing on the deformation of the droplet: too low a value corresponds to weak elasticity and may elicit a Newtonian-like response from the droplet, while too high a value might make the elastic contribution so strong that it inhibits any inertial deformation of the droplet. Therefore, the value of $\Pi$ that is chosen for the current setup is decided after some trial-and-error iterations to best showcase the viscoelastic response of the liquid. The viscoelastic filament morphologies observed in figure \ref{fig:fig-6}b for $t >$ 100 are well-described by the schematic representation in figure \ref{fig:fig-4}d-iii. In other words, the viscoelastic filament (or bag) exhibits a much more uniform thinning than a Newtonian filament (or bag). This indeed leads to a delayed fragmentation of the bags~(increased lifetimes), and consequently longer bags---observations that are in complete agreement with our experimental findings shown in figure \ref{fig:fig-3}. 

We further quantify this thinning behavior by calculating the mean-squared deviation of the filament thickness, $h$, about its minimum value, $h_{\min}$, expressed as
\begin{align}
	\hat{\sigma}^{2} &= \frac{1}{N} \sum_{i = 1}^{N} \left( h_{i} - h_{\min} \right)^{2} \, ,
	\label{eq:eq-variance}
\end{align}
where $N$ is the total number of points where the thickness of the filament is sampled (for instance, the black segments in figure \ref{fig:fig-7}a). The temporal variation of this mean-squared deviation, $\hat{\sigma}^{2}$, is shown in figure \ref{fig:fig-7}b for different Weissenberg number $\mathrm{Wi}$ values (while the elastoinertial number $\Pi$ = 0.001 is kept constant). A larger $\mathrm{Wi}$ corresponds to a stronger viscoelastic response as the polymers remain highly-stretched within a time duration of the inertial time scale $\tau_{\mathrm{d}}$, i.e. the liquid retains its memory of its previous deformation states to a greater extent. Hence, as $\mathrm{Wi}$ increases, $\hat{\sigma}^{2}$ decreases, indicating that the thinning of the bags occurs in a more uniform manner. This, in turn, leads to a delayed breakup, as inferred from the time instants corresponding to the last data point for each $\mathrm{Wi}$ in figure \ref{fig:fig-7}b (the temporal tracking of $\hat{\sigma}^{2}$ is halted when the bag breaks up). This further confirms mechanisms underlying the experimentally-observed trends for viscoelastic bags in figure \ref{fig:fig-3}: a more uniform thinning (see figure \ref{fig:fig-4}c-iii) leads to thinner bags, which, in turn, are larger in size and have a longer lifetime. Consequently, these thinner filaments, upon break up, generate smaller droplets, as also observed in our experiments (see figure \ref{fig:fig-4}).

\section{Conclusions and outlook} \label{sec:conclusions}

In summary, we have studied the shear-induced fragmentation of an idealized mucosalivary-mimetic fluid in a closed geometry, which is analogous to the process of bioaerosol generation in the human trachea during violent respiratory events such as coughing and sneezing. Through experiments and numerical simulations, we have investigated the effect of viscoelasticity of the mucosalivary-mimetic fluid on the fragmentation behavior and the resulting droplet size distribution. While both Newtonian and viscoelastic liquid films fragment via the bag breakup mode, viscoelastic liquids tend to sustain larger inflated bag structures, with an increased lifetime, prior to rupture. This, in turn, leads to the generation of smaller droplets during the unstable retraction of the liquid rims when the bags eventually rupture. The reduction in droplet sizes is attributed to the thinner bags of comparatively more uniform thickness formed in the viscoelastic cases, which is also supported by our numerical simulations. 

The observed reduction in droplet size with increasing viscoelasticity has significant implications in the context of airborne pathogen transmission and its control. Smaller droplets with a diameter of $\mathcal{O}$(10 \SI{}{\text{\micro} \text{m}}) can persist in the turbulent exhalation puff, and stay afloat for an extended period of time, as well as being transported over larger distances \citep{ng-2021-prf, wang-2021-pnas}. This, in turn, affects conventional mitigation measures such as ventilation and social distancing. Therefore, infection control strategies targeting both the generation and transmission of bioaerosols must take into account the role played by viscoelasticity of the mucosalivary fluid in droplet generation. The findings of the present work represents a small-yet-significant step in that direction. 

\section*{Acknowledgements}

We acknowledge funding from the Max Planck Center Twente for Complex Fluids, the Netherlands Organization for Scientific Research (NWO) through the MIST project with project number P20-35 of the Perspectief research programme, the European Research Council (ERC) through Advanced Grant numbers 740479 and 883849, and the Netherlands Organization for Health Research and Development (ZonMW) through grant number 10430012010022. The authors are grateful for the technical assistance from Gert-Wim Bruggert during the fabrication of the experimental setup. The authors thank Lydia Bourouiba, Alvaro Marin, Gareth McKinley, and Vatsal Sanjay for insightful discussions. 

\bibliography{cough-machine.bib}


\end{document}